\documentclass[11pt]{article}

\usepackage{amsmath}  
\usepackage{amssymb}
\usepackage{amsthm}
\usepackage{fancybox}
\usepackage{times}

\theoremstyle{definition}
\newtheorem{definition}{Definition}
\theoremstyle{theorem}
\newtheorem{theorem}{Theorem}
\newtheorem{lemma}{Lemma}

\def\bbbone{{\mathchoice {\rm 1\mskip-4mu l} {\rm 1\mskip-4mu l}
{\rm 1\mskip-4.5mu l} {\rm 1\mskip-5mu l}}}
\newcommand{\sss}{\scriptscriptstyle}

\newcommand{\integers}{\mathbb{Z}}
\newcommand{\reals}{\mathbb{R}}
\newcommand{\complex}{\mathbb{C}}
\newcommand{\field}{\mathbb{F}}

\newcommand{\sym}{\mbox{SYM}}
\newcommand{\qinit}{q_{\sss 0}}
\newcommand{\pinit}{p_{\sss 0}}

\newcommand{\Insert}{\vdash}
\def\Hcal{\mathcal{H}}
\def\Hint{\mathfrak{H}}
\def\Dcal{\mathcal{D}}
\def\Ccal{\mathcal{C}}
\def\Qcal{\mathcal{Q}}
\def\Xcal{\mathcal{X}}
\def\Ical{\mathcal{I}}
\def\kernel{{\rm kernel}}
\def\Gau{{\rm Gau}}
\def\Ophys{\mathcal{O}_{\!\scriptscriptstyle \rm phys}}
\def\Fun{\mathcal{F}}
\def\Funinf{\mathcal{F}_{\!\!\scriptscriptstyle\infty}}
\def\Funpol{\mathcal{F}_{\!\!\scriptscriptstyle\rm pol}}
\def\Funpolone{\mathcal{F}_{\!\!\scriptscriptstyle\rm pol(1)}}
\def\Funpoltwo{\mathcal{F}_{\!\!\scriptscriptstyle\rm pol(2)}}
\def\Funirr{\mathcal{F}_{\!\!\scriptscriptstyle\rm irr}}
\def\Funquant{\mathcal{F}_{\!\!\scriptscriptstyle\rm quant}}
\def\Funpolinfone{\mathcal{F}_{\!\!\scriptscriptstyle\rm pol(\infty,1)}}
\def\i{{\rm i}}
\def\sss{\scriptscriptstyle}
\begin{document}

\title{That Strange Procedure Called Quantisation}

\author{Domenico Giulini            \\
        University of Freiburg      \\
        Department of Physics       \\
        Hermann-Herder-Strasse 3    \\
        D-79104 Freiburg,           \\
        Germany}
\date{}

\maketitle
\begin{abstract}
\noindent
This is a pedagogical and (almost) self-contained introduction into the 
Theorem of Groenewold and van\,Howe, which states that a naive transcription 
of Dirac's quantisation rules cannot work. Some related issues in 
quantisation theory are also discussed. First-class constrained systems 
a briefly described in a slightly more `global' fashion.\footnote{  
This is the written version of a lecture delivered at the school 
``Aspects of Quantum Gravity - From Theory to Experimental Search'', 
held between February \,26. -- March\,1. in 2002 at the 
Physics Center of the German Physical Society in Bad Honnef (Germany);
see $\langle$http://kaluza.physik.uni-konstanz.de/CL/QG/$\rangle$. 
To be published in the proceedings, edited by D. Giulini, C. Kiefer, 
and C. L\"ammerzahl, at Springer Verlag (Berlin, 2003).} 
 
\end{abstract}

\maketitle              
\section{Introduction and Motivation}
\label{sec:introduction}
In my contribution I wish to concentrate on some fundamental issues 
concerning the notion of \emph{quantisation}. Nothing of what I will say 
is new or surprising to the experts. My intention is rather 
a pedagogical one: to acquaint the non-experts with some of the  
basic structural results in quantisation theory, which I feel should be 
known to anybody who intends to `quantise' something. A central result 
is the theorem of Groenewold and van\,Hove, which is primarily a no-go 
result, stating that the most straightforward axiomatisation of Dirac's 
informally presented `canonical' quantisation rules runs into contradictions 
and therefore has to be relaxed. The constructive value of this theorem 
lies in the fact that its proof makes definite suggestions for such 
relaxations. This helps to sharpen ones expectations on the quantisation 
concept in general, which is particularly important for Quantum Gravity
since here sources for direct physical input are rather scarce. 
Expectations on what Quantum Gravity will finally turn out to be 
are still diverse, though more precise pictures now definitely emerge 
within the individual approaches, as you will hopefully be convinced in
the other lectures~\cite{Loll:2003,Mohaupt:2003,Thiemann:2003}, so that 
reliable statements about similarities and differences on various points 
can now be made. The present contribution deliberately takes focus on 
a very particular and seemingly formal point, in order to exemplify 
in a controllable setting the care needed in formulating `rules' for 
`quantisation'. At the end I will also briefly consider constrained 
systems from a slightly more `global' point of view. Two appendices 
provide some technical aspects.  
 
How do you recognize \emph{quantum} theories and what structural elements
distinguish them from so-called classical ones? If someone laid down,
in mathematical terms, a theory of `something' before you, what 
features would you check in order to answer this question? Or would 
you rather maintain that this question does not make good sense to 
begin with? Strangely enough, even though quantum theories are not only 
known to be the most successful but also believed to be the most 
fundamental theories of physics, there seems to be no unanimously 
accepted answer to any of these questions. So far a working hypothesis 
has been to \emph{define} quantum theories as the results of some 
`quantisation procedures' after their application to classical theories. 
One says that the classical theory (of `something') `gets quantised' 
and that the result is the quantum theory (of that `something'). 
This is certainly the way we traditionally understand Quantum Mechanics and 
also a substantial part of Quantum Field Theory (for more discussion 
on this point, that also covers interesting technical issues, I 
recommend \cite{Isham-LesHouches:1984}). As an exception---to a 
certain degree---I would list Local Quantum Field Theory~\cite{Haag:1996}, 
which axiomatically starts with a general kinematical framework for 
Poincar\'e invariant quantum field theories without any a priori 
reference to classical theories. Although this can now be generalised 
to curved spacetimes, it does not seem possible to eliminate the need 
of some such fixed (i.e. non-dynamical) background. Hence this approach 
does not seem to be able to apply to background independent dynamical 
fields, like gravity.

The generally accepted quantisation procedures I have in mind here can 
be roughly divided into three groups, with various interrelations:
\begin{itemize}
\item
Hilbert-space based methods, like the standard canonical quantisation
programme,
\item
algebraic methods based on the notion on observables, like $\star$-product 
quantisation or $C^*$-algebra methods,
\item 
path integral methods.  
\end{itemize}
Given the success of Quantum Mechanics (QM) it was historically, 
and still is, more than justified to take it as paradigm for all other 
quantum theories (modulo extra technical inputs one needs to handle 
infinitely many degrees of freedom). Let us therefore take a look at 
QM and see how quantisation may, or may not, be understood.  
In doing this, I will exclusively focus on the traditional `canonical' 
approaches to quantisation. 

\section{Canonical Quantisation}
Historically the rules for `canonical quantisation' where first spelled out 
by Dirac in his famous book on QM~\cite{Dirac:1958}. His followers 
sometimes bluntly restated these rules by the symbolic line, 
\begin{equation}
\label{eq:naive-Dirac-quantisation}
\{\cdot\,,\,\cdot\}\mapsto\tfrac{-\i}{\hbar}[\cdot\,,\cdot]\,,
\end{equation}
which is to be read as follows: map each classical observable 
(function on phase space) $f$ to an operator $\hat f$ in a Hilbert 
space $\Hcal$ (typically $L^2(Q,d\mu)$, where $Q$ is the classical 
configuration space and $d\mu$ the measure that derives from the 
Riemannian metric thereon defined by the kinetic energy), in such a 
way that the Poisson bracket of two observables is mapped to 
$-\i/\hbar$ times the commutator of the corresponding operators, 
i.e, $\widehat{\{f_1,f_2\}}=\tfrac{-\i}{\hbar}[{\hat f}_1,{\hat f}_2]$.
This is also facetiously known as `quantisation by hatting'. 
But actually Dirac was more careful; he wrote \cite{Dirac:1958} 
(my emphasis; P.B. denotes `Poisson Brackets')

\begin{center}
\setlength{\fboxsep}{3.0mm}
\setlength{\fboxrule}{1.5pt}
\fbox{\parbox{10cm}{
`The strong analogy between quantum P.B. [i.e. commutators]
and classical P.B. leads us to make the assumption that the 
quantum P.B., \emph{or at any rate the simpler ones of them,} 
have the same values as the corresponding classical P.B.s.'

\medskip
\hfill 
{\sc Paul Dirac, 1930}
}}
\end{center} 
Clearly these words demand a specific interpretation before 
they can be called a (well defined) quantisation programme.

\subsection{The classical stage}
\label{sec:the-classical-stage}
Associated to a classical Hamiltonian dynamical system of $n$ 
degrees of freedom is a $2n$-dimensional manifold, $P$, the 
space of states or `phase space' (sometimes identified with 
the space of solutions to Hamilton's equations, if the latter
pose a well defined initial-value problem). Usually---but not 
always---it comes equipped with a preferred set of $2n$ 
functions, $(q^i,p_i)$, $i=1\cdots n$, called coordinates
and momenta respectively. In addition, there is a 
differential-geometric structure on $P$, called $\emph{Poisson Bracket}$, 
which gives a suitable subspace $\Fun\subseteq C^{\infty}(P)$ of the 
space of real-valued, infinitely differentiable functions the structure 
of a Lie algebra. See Appendix\,1 for more information on the geometric 
structures of classical phase space and Appendix\,2 for the general 
definition of a Lie algebra. Exactly what subspace is `suitable' 
depends of the situation at hand and will be left open for the time being. 
In any case, the Poisson Bracket is a map 
\begin{equation}
\label{eq:PB-map}
\{\cdot,\cdot\}:\ \Fun\times\Fun\rightarrow\Fun\,,
\end{equation}
which satisfies the following conditions for all $f,g,h\in\Fun(P)$
and $\lambda\in\reals$ (which make it precisely a real Lie algebra):
\begin{alignat}{2}
\label{eq:Lie-def-1}
&\{f,g\}=-\{g,f\}
\qquad &&\mbox{antisymmetry}\,,\\
\label{eq:Lie-def-2}
&\{f,g+\lambda h\}=\{f,g\}+\lambda\{f,h\}
\qquad &&\mbox{linearity}\,,\\
\label{eq:Lie-def-3}
&\{f,\{g,h\}\}+\{g,\{h,f\}\}+\{h,\{f,g\}\}=0
\qquad &&\mbox{Jacobi identity}\,.
\end{alignat}
In the special coordinates $(q^i,p_i)$ it takes the explicit form
(cf. Appendix\,1)
\begin{equation}
\label{eq:PB-def}
\{f,g\}:=\sum_{i=1}^n
\left(
\frac{\partial f}{\partial q^i}\frac{\partial g}{\partial p_i}-
\frac{\partial f}{\partial p_i}\frac{\partial g}{\partial q^i}
\right)\,.  
\end{equation}

Independently of the existence of a Poisson Bracket, the space 
$\Fun$ is a commutative and associative algebra under the operation
of pointwise multiplication:
\begin{equation}
\label{eq:ass-alg1}
(f\cdot g)(x):=f(x)g(x)\,.  
\end{equation}
This means that the multiplication operation is also a map 
$\Fun\times\Fun\rightarrow\Fun$ (simply denoted by `$\cdot$')
which satisfies the following conditions for all $f,g,h\in\Fun$
and $\lambda\in\reals$:

\begin{alignat}{2}
\label{eq:ass-def-1}
&f\cdot g = g\cdot f
\qquad &&\mbox{commutativity}\,,\\
\label{eq:ass-def-2}
&f\cdot(g+\lambda\, h)=f\cdot g+\lambda\,f\cdot h
\qquad &&\mbox{linearity}\,,\\
\label{eq:ass-def-3}
&f\cdot(g\cdot h)=(f\cdot g)\cdot h
\qquad &&\mbox{associativity}\,.
\end{alignat}

The two structures are intertwined by the following condition, which 
expresses the fact that each map $D_f:\Fun\rightarrow\Fun$, 
$g\mapsto D_f(g):=\{f,g\}$, is a derivation of the associative algebra 
for each $f\in\Fun$:
\begin{equation}
\label{eq:derivation}
\{f,g\cdot h\}=\{f,g\}\cdot h+g\cdot\{f,h\}\,.
\end{equation}
The Jacobi identity now implies that ($\circ$ denotes composition)
$D_f\circ D_g-D_g\circ D_f=D_{\{f,g\}}$.\footnote{This can be 
expressed by saying that the assignment $f\mapsto D_f$ is a Lie 
homomorphism from the Lie algebra $\Fun$ to the Lie algebra of 
derivations on $\Fun$. Note that the derivations form an
associative algebra when multiplication is defined to be 
composition, and hence also a Lie algebra when the Lie product 
is defined to be the commutator.} Taken all this together 
this makes $\Fun$ into a Poisson algebra, whose abstract 
definition is as follows:
\begin{definition}
A \emph{Poisson algebra} is a vector space $V$ with two maps 
$V\times V\rightarrow V$, denoted by `$\{{},{}\}$' and `$\cdot$', 
which turn $V$ into a Lie algebra (defined by 
(\ref{eq:Lie-def-1}-\ref{eq:Lie-def-3})) and 
a commutative and associative algebra (defined by 
(\ref{eq:ass-def-1}-\ref{eq:ass-def-3})) respectively, such that 
(\ref{eq:derivation}) holds. 
\end{definition}

Simply writing the symbol $\Fun$  now becomes ambiguous since 
it does not indicate which of these different structures we wish 
to be implicitly understood. I shall use the convention 
to let `$+$' indicate the vector-space structure, $(+,\{,\})$ the 
Lie-algebra structure, $(+,\cdot)$ the associative structure, and 
$(+,\{{},{}\},\cdot)$ the Poisson structure. To avoid confusion I
will then sometimes write:
\begin{alignat}{2}
\label{eq:Fun-defs-1}
&\Fun                   &&\quad\mbox{for the set}\,,\\
\label{eq:Fun-defs-2}
&\Fun(+,\{{},{}\})      &&\quad\mbox{for the Lie algebra}\,,\\
\label{eq:Fun-defs-3}
&\Fun(+,\cdot)          &&\quad\mbox{for the associative algebra}\,,\\    
\label{eq:Fun-defs-4}
&\Fun(+,{\{},{}\},\cdot)&&\quad\mbox{for the Poisson algebra}\,,
\end{alignat}
formed by our subset of functions from $C^{\infty}(P)$. Sometimes 
I will indicate the subset of functions by a subscript on $\Fun$. 
For example, I will mostly restrict $P$ to be $\reals^{2n}$ with 
coordinates $(q^i,p_i)$. It then makes sense to restrict to functions 
which are polynomials in these coordinates.\footnote{Recall that you need 
an affine structure on a space in order to give meaning to the term 
`polynomial functions'.} Then the following subspaces will turn out 
to be important in the sequel:
\begin{alignat}{3}
\label{eq:Fun-subs-1}
&\Funinf       &&:&&\quad\mbox{$C^{\infty}$-functions,}              \\
&\Funpol       &&:&&\quad\mbox{polynomials in $q$'s and $p$'s,}      \\
&\Funpolone    &&:&&\quad\mbox{polynomials of at most first order,}  \\
&\Funpoltwo    &&:&&\quad\mbox{polynomials of at most second order,} \\
&\Funpolinfone &&:&&\quad\mbox{polynomials of at most first order in 
                               the $p$'s}\nonumber \\
&\phantom{\Funpolinfone}&&\phantom{:}&&\quad\mbox{
               whose coefficients are polynomials in the $q$'s.} \quad
\end{alignat}
An otherwise unrestricted polynomial dependence is clearly preserved 
under addition, scalar multiplication, multiplication of functions, 
and also taking the Poisson Bracket (\ref{eq:PB-def}). 
Hence $\Funpol$ forms a  Poisson subalgebra. This is not true for the 
other subspaces listed above, which still form Lie subalgebras but not 
associative algebras.

\subsection{Defining `canonical quantisation'}
\label{sec:can-quant-def}
Roughly speaking, Dirac's approach to quantisation consists in 
mapping certain functions on $P$ to the set $\sym(\Hcal)$ 
of symmetric operators (sometimes called `formally self adjoint') 
on a Hilbert space $\Hcal$. Suppose these operators have a common 
invariant dense domain  $\Dcal\subset\Hcal$ (typically the `Schwarz 
space'), then it makes sense to freely multiply them. This generates 
an associative algebra of operators (which clearly now also contains 
non-symmetric ones) defined on $\Dcal$. Note that every associative 
algebra is automatically a Lie algebra by defining the Lie product 
proportional to the commutator
(cf. Appendix\,2):
\begin{equation}
\label{eq:Lie-commutator}
[X,Y]:=X\cdot Y-Y\cdot X\,.  
\end{equation}
Since the commutator of two symmetric operators is antisymmetric, we obtain 
a Lie-algebra structure on the real vector space of symmetric operators 
with invariant dense domain $\Dcal$ by defining the Lie product as 
imaginary multiple of the commutator; this I will write 
as $\tfrac{1}{\i\hbar}[\cdot,\cdot]$ where $\hbar$ is a real (dimensionful)
constant, eventually to be identified with Planck's constant divided by 
$2\pi$. 

Note that I deliberately did \emph{not} state that classical observables 
should be mapped to \emph{self adjoint} operators. Instead I only required 
the operators to be symmetric, which is a weaker requirement. This 
important distinction (see e.g. 
\cite{Reed-Simon-1:1972}) is made for the following reason (see e.g. 
sect.\,VIII in \cite{Reed-Simon-1:1972} for the mathematical distinction): 
let $\hat f$ be the operator corresponding to the phase-space function 
$f$. If $\hat f$ were self adjoint, then the quantum flow 
$U(t)=\exp(it\hat f)$ existed for all $t\in\reals$, even if the classical 
Hamiltonian vector field for $f$ is incomplete (cf. Appendix\,1) so that 
the classical flow does not exist for all flow parameters in $\reals$. 
Hence self adjointness seems too strong a requirement for such $f$ 
whose classical flow is incomplete (which is the generic situation). 
Therefore one generally only requires the operators to be symmetric and 
strengthens this explicitly for those $f$ whose classical flow is 
complete (see below). 

A first attempt to mathematically define Dirac's quantisation 
strategy could now consist in the following: find a `suitable' 
Lie homomorphism $\Qcal$ from a `suitable' Lie subalgebra 
$\Fun'\subset\Fun(+,\{{},{}\})$ to the Lie algebra $\sym(\Hcal)$ 
of symmetric operators on a Hilbert space $\Hcal$ with some common dense 
domain $\Dcal\subset\Hcal$. The map $\Qcal$ will be called the 
\emph{quantisation map}. Note that this map is a priori not required in 
any way to preserve the associative structure, i.e. no statement is 
made to the effect that $\Qcal(f\cdot g)=\Qcal(f)\cdot\Qcal(g)$, or 
similar. 

To be mathematically precise, we still need to interpret the word 
`suitable' which occurred twice in the above statement. For this
we consider the following \emph{test case}, which at first sight 
appears to be sufficiently general and sufficiently precise to 
be able to incorporate Dirac's ideas in a well defined manner:
\begin{enumerate}
\item
We restrict the Lie algebra of $C^{\infty}$-Functions on $P$ to 
polynomials in $(q^i,p_i)$, i.e. we consider 
$\Funpol(+,\{{},{}\})$.
\item
As Hilbert space of states, $\Hcal$,  we consider the space of 
square-integrable functions $\reals^n\rightarrow\Hint$, where $\Hint$
is a \emph{finite} dimensional Hilbert space which may account 
for internal degrees of freedom, like spin. $\reals^n$ should 
be thought of as `half' of phase space, or more precisely the 
configuration space coordinatised by the set $\{q^1,\cdots,q^n\}$. 
For integration we take the Lebesgue measure $d^nq$. 
\item
There exists a map 
$\Qcal:\Funpol\rightarrow \sym(\Hcal,\Dcal)$ into the set of symmetric
operators on $\Hcal$ with common invariant dense domain $\Dcal$. 
(When convenient we also write $\hat f$ instead of $\Qcal(f)$.) 
This map has the property that whenever $f\in\Funpol$ has a complete 
Hamiltonian vector field the operator $Q(f)$ is in fact (essentially)
self adjoint.\footnote{We remark that the subset of functions whose 
flows are complete do not form a Lie subalgebra; hence it would not 
make sense to just restrict to them.} 
\item
$\Qcal$ is linear: 
\begin{equation}
\label{eq:Q-linear}
\Qcal(f+\lambda\, g)=\Qcal(f)+\lambda\,\Qcal(g)\,.
\end{equation}
\item
$\Qcal$ intertwines the Lie structure on $\Funpol(+,\{{},{}\})$ and 
the Lie structure given by $\tfrac{1}{\i\hbar}[{},{}]$ on 
$\sym(\Hcal,\Dcal)$:
\begin{equation}
\label{eq:Q-Lie-preserving}
\Qcal(\{f,g\}))= \tfrac{1}{\i\hbar}[\Qcal(f),\Qcal(g)])\,.
\end{equation}
Here $\hbar$ is a constant whose physical dimension is that of $p\cdot q$
(i.e. an action) which accounts for the intrinsic dimension of $\{{},{}\}$ 
acquired through the differentiations (cf. (\ref{eq:PB-def})). 
Note again that the imaginary unit is necessary to obtain a Lie structure 
on the subset of symmetric operators. 
 \item
Let $1$ also denote the constant function with value $1$ on $P$ and 
$\bbbone$ the unit operator; then 
\begin{equation}
\label{eq:Q-unit-preserving}
\Qcal(1)=\bbbone\,.
\end{equation}
\item
The quantisation map $\Qcal$ is consistent with 
Schr\"odinger quantisation:
\begin{alignat}{1}
\label{eq:Q-Schrodinger-consistency1}
(\Qcal(q^i)\psi)(q)&\,=\,q^i\psi(q)\,,\\
\label{eq:Q-Schrodinger-consistency2}
(\Qcal(p_i)\psi)(q)&\,=\,-\i\hbar\partial_{q^i}\psi(q)\,.
\end{alignat}
\end{enumerate}

One might wonder what is actually implied by the last condition and 
whether it is not unnecessarily restrictive. This is clarified by 
the theorem of \emph{Stone} and \emph{von Neumann} (see e.g. 
\cite{Abraham-Marsden:1978}), which says that if the $2n$ operators 
$\Qcal(q^i)$ and $\Qcal(p_i)$ are represented \emph{irreducibly up 
to finite multiplicity} (to allow for finitely many internal 
quantum numbers) and satisfy the required commutation relations, 
then their representation is unitarily equivalent to the Schr\"odinger 
representation given above. In other words, points 2.) and 7.) above 
are equivalent to, and could therefore be replaced by, the following 
requirement:
\begin{enumerate}
\item[7'.]
The $2n$ operators $\Qcal(q^i), \Qcal(p_i)$ act irreducibly up to 
at most finite multiplicity on $\Hcal$.
\end{enumerate}

Finally there is a technical point to be taken care of. Note that 
the commutator on the right hand side of 
(\ref{eq:Q-Lie-preserving})---and hence the whole equation---only 
makes sense on the subset $\Dcal\subseteq\Hcal$. This becomes 
important if one deduces from (\ref{eq:Q-linear}) and  
(\ref{eq:Q-Lie-preserving}) that 
\begin{equation}
\label{Q:commuting-obs}
\{f,g\}=0\Rightarrow [\Qcal(f),\Qcal(g)]=0\,,
\end{equation}
i.e. that $\Qcal(f)$ and $\Qcal(g)$ commute \emph{on} $\Dcal$. 
Suppose that the Hamiltonian vector fields of $f$ and $g$ are 
complete so that $Q(f)$ and $Q(g)$ are self adjoint. Then commutativity 
on $\Dcal$ does \emph{not} imply that $\Qcal(f)$ and $\Qcal(g)$ commute 
in the usual (strong) sense of commutativity of self-adjoint operators, 
namely that all their spectral projectors mutually commute 
(compare \cite{Reed-Simon-1:1972}, p.\,271). This we pose as an extra 
condition:
\begin{enumerate}
\item[8.]
If $f,g$ have complete Hamiltonian vector fields and $\{f,g\}=0$; 
then $\Qcal(f)$ commutes with $\Qcal(g)$ in the strong sense, i.e. 
their families of spectral projectors commute. 
\end{enumerate}
This extra condition will facilitate the technical presentation 
of the following arguments, but we remark that it can be dispensed 
with \cite{Gotay:1999}.  

\subsection{The theorem of Groenewold and van Howe}
\label{sec:GvH-Theorem}
In a series of papers Groenewold \cite{Groenewold:1946} 
and van Hove \cite{VanHowe:1951-a,VanHowe:1951-b}
showed that a canonical quantisation satisfying requirements 
1.--8. does \emph{not} exist. The proof is instructive and 
therefore we shall present it in detail. For logical clarity 
it is advantageous to divide it into two parts:

\smallskip
\noindent
{\bf Part\,1} shows the following `squaring laws':
\begin{alignat}{2}
\label{eq:squaring-law-1}
&\Qcal(q^2)\,&&=\,[\Qcal(q)]^2\,,\\
\label{eq:squaring-law-2} 
&\Qcal(p^2)\,&&=\,[\Qcal(p)]^2\,,\\
\label{eq:squaring-law-3}
&\Qcal(qp) \,&&=\,\tfrac{1}{2}[\Qcal(q)\Qcal(p)+\Qcal(p)\Qcal(q)]\,.
\end{alignat}
Next to elementary manipulations the proof of part\,1 uses a 
result concerning the Lie algebra $sl(2,\reals)$, which we shall 
prove in Appendix\,2.
Note that in the canonical approach as formulated here \emph{no}
initial assumption whatsoever was made concerning the preservation 
of the associative structure. Points 4. and 5. only required the 
Lie structure to the preserved. The importance of part\,1 is to 
show that such a partial preservation of the associative structure 
can actually be derived. It will appear later 
(cf. Sect.\,\ref{sec:irreducibility}) that this consequence could 
not have been drawn without the irreducibility requirement~7').   

\smallskip
\noindent   
{\bf Part\,2} shows that the squaring laws lead to a contradiction 
to (\ref{eq:Q-Lie-preserving}) on the level of higher than second-order 
polynomials. 

\smallskip
\noindent
Let us now turn to the proofs. To save notation we write 
$\hat f$ instead of $\Qcal(f)$. Also, we restrict attention to $n=1$, i.e. 
we have one $q$ and one $p$ coordinate on the two dimensional phase space 
$\reals^2$. In what follows, essential use is repeatedly made of 
condition\,8 in the following form: assume $\{f,q\}=0$ then 
(\ref{eq:Q-Lie-preserving}) and condition\,8 require that $\hat f$ 
(strongly) commutes with $\hat q$, which in the Schr\"odinger 
representation implies that $\hat f$ has the form 
$(\hat f\psi)(q)=A(q)\psi(q)$, where $A(q)$ is a Hermitean operator 
(matrix) in the finite dimensional (internal) Hilbert space $\Hint$.    

\bigskip\noindent
{\bf Proof of part\,1} We shall present the argument in 7 small steps.
Note that throughout we work in the Schr\"odinger representation.

\begin{itemize}
\item[i)]
Calculate $\widehat{q^2}$: we have $\{q^2,q\}=0$, hence 
$\widehat{q^2}=A(q)$. Applying (\ref{eq:Q-Lie-preserving}) 
and (\ref{eq:Q-Schrodinger-consistency1}) to $\{p,q^2\}=-2q$ 
gives 
$\tfrac{1}{\i\hbar}[\hat p,\widehat{q^2}]=-2\hat q$ and hence
$A'(q)=2q$ (here we suppress to write an explicit $\bbbone$ for the 
unit operator in $\Hint$), so that 
\begin{equation}
\label{eq:GvH-proof-1}
\widehat{q^2}={\hat q}^2-2\mathfrak{e}_-\,,   
\end{equation}
where $\mathfrak{e}_-$ is a constant (i.e. $q$ independent) Hermitean 
matrix in $\Hint$.
\item[ii)]
Calculate $\widehat{p^2}$: this is easily obtained by just
Fourier transforming the case just done. Hence 
\begin{equation}
\label{eq:GvH-proof-2}
\widehat{p^2}={\hat p}^2+2\mathfrak{e}_+\,,   
\end{equation} 
where $\mathfrak{e}_+$ is a constant Hermitean matrix in $\Hint$
(here, as in (\ref{eq:GvH-proof-1}), the conventional factor of 2 and the 
signs are chosen for later convenience).
\item[iii)]
Calculate $\widehat{qp}$: We apply (\ref{eq:Q-Lie-preserving}) to
$4qp=\{q^2,p^2\}$ and insert the results (\ref{eq:GvH-proof-1}) and 
(\ref{eq:GvH-proof-2}):
\begin{equation}
\label{eq:GvH-proof-3}
\widehat{qp}=\tfrac{1}{4\i\hbar}[\widehat{q^2},\widehat{p^2}]
             =\tfrac{1}{4\i\hbar}[\hat q^2,\hat p^2]
             -\tfrac{1}{\i\hbar}[\mathfrak{e}_-,\mathfrak{e}_+]
             =\tfrac{1}{2}(\hat q\hat p+\hat p\hat q)+\mathfrak{h}\,,
\end{equation}
where
\begin{equation}
\label{eq:GvH-proof-4}
\mathfrak{h}:=\tfrac{1}{\i\hbar}[\mathfrak{e}_+,\mathfrak{e}_-]\,.
\end{equation}
In the last step of (\ref{eq:GvH-proof-3}) we iteratively used the 
general rule 
\begin{equation}
\label{eq:GvH-proof-derivationrule}
[A,BC]=[A,B]C+B[A,C]\,.  
\end{equation}
\item[iv)]
Next consider the quantities 
\begin{alignat}{2}
\label{eq:GvH-proof-5} 
& h\,&&:=\,\tfrac{1}{2}(\hat q\hat p+\hat p\hat q)\,,\\
\label{eq:GvH-proof-6} 
& e_+\,&&:=\,\tfrac{1}{2}{\hat p}^2               \,,\\
\label{eq:GvH-proof-7} 
& e_-\,&&:=\,-\tfrac{1}{2}{\hat q}^2              \,.
\end{alignat}
By straightforward iterative applications of 
(\ref{eq:GvH-proof-derivationrule}) short computations yield
\begin{equation}
\label{eq:GvH-proof-7b} 
\tfrac{1}{\i\hbar}[e_+,e_-]= h\,,\quad
\tfrac{1}{\i\hbar}[h,e_{\pm}]=\pm 2\, e_{\pm}\,,
\end{equation}
which show that $e_\pm,h$ furnish a representation of the Lie algebra
of $sl(2,\reals)$ of real traceless $2\times 2$ matrices 
(see Appendix\,2 for details).  
\item[v)]
On the other hand, defining 
\begin{alignat}{2}
\label{eq:GvH-proof-8} 
& H\,&&:=\,\widehat{qp}                 \,,\\
\label{eq:GvH-proof-9} 
& E_+\,&&:=\,\tfrac{1}{2}\widehat{p^2}  \,,\\
\label{eq:GvH-proof-10} 
& E_-\,&&:=\,-\tfrac{1}{2}\widehat{q^2} \,,
\end{alignat}
we can directly use (\ref{eq:Q-Lie-preserving}) to calculate their 
Lie brackets. This shows that they also satisfy the 
$sl(2,\reals)$ algebra:
\begin{equation}
\label{eq:GvH-proof-11} 
\tfrac{1}{\i\hbar}[E_+,E_-]= H\,,\quad
\tfrac{1}{\i\hbar}[H,E_{\pm}]=\pm 2\,E_{\pm}\,.
\end{equation}
\item[vi)]
Inserting into (\ref{eq:GvH-proof-11}) the results 
(\ref{eq:GvH-proof-1})~(\ref{eq:GvH-proof-2})~(\ref{eq:GvH-proof-3})
now implies that the Hermitean matrices $\mathfrak{e}_{\pm},\mathfrak{h}$
too satisfy the $sl(2,\reals)$ algebra:
\begin{equation}
\label{eq:GvH-proof-12} 
\tfrac{1}{\i\hbar}[\mathfrak{e}_+,\mathfrak{e}_-]=\mathfrak{h}\,,\quad
\tfrac{1}{\i\hbar}[\mathfrak{h},\mathfrak{e}_{\pm}]
=\pm 2\,\mathfrak{e}_{\pm}\,.
\end{equation}
\item[vii)]
Finally we invoke the following result from Appendix\,2:
\begin{lemma}
\label{lemma:sl2}
Let $A,B_+,B_-$ be finite dimensional anti-Hermitean matrices which 
satisfy $A=[B_+,B_-]$ and $[A,B_\pm]=\pm 2B_\pm$, then $A=B_\pm=0$.
\end{lemma}
Applying this to our case by setting $A=\tfrac{1}{\i\hbar}\mathfrak{h}$ 
and $B_\pm=\tfrac{1}{\i\hbar}\mathfrak{e}_\pm$ implies
\begin{equation}
\label{eq:eq:GvH-proof-13}
\mathfrak{e}_\pm=0=\mathfrak{h}\,.
\end{equation} 
Inserting this into (\ref{eq:GvH-proof-1}-\ref{eq:GvH-proof-3}) 
yields (\ref{eq:squaring-law-1}-\ref{eq:squaring-law-3}) 
respectively. This ends the proof of part\,1.
\end{itemize}

\bigskip\noindent
{\bf Proof of part\,2}

\smallskip\noindent
Following \cite{Gotay:1999}, we first observe that the 
statements~(\ref{eq:squaring-law-1}-\ref{eq:squaring-law-3}) 
can actually be generalised: Let $P$ be any real polynomial, 
then 

\begin{alignat}{2}
\label{eq:gen-squaring-law-1}
&\widehat{P(q)}&&\,=\,P(\hat q)\,,\\
\label{eq:gen-squaring-law-2}
&\widehat{P(p)}&&\,=\,P(\hat p)\,,\\
\label{eq:gen-squaring-law-3}
&\widehat{P(q)p}&&\,=\,\tfrac{1}{2}(P(\hat q)\hat p+\hat pP(\hat q))\,,\\
\label{eq:gen-squaring-law-4}
&\widehat{P(p)q}&&\,=\,\tfrac{1}{2}(P(\hat p)\hat q+\hat qP(\hat p))\,.
\end{alignat}
To complete the proof of part\,2 it is sufficient to prove 
(\ref{eq:gen-squaring-law-1}) and (\ref{eq:gen-squaring-law-2})
for $P(x)=x^3$, and (\ref{eq:gen-squaring-law-3}) and 
(\ref{eq:gen-squaring-law-4}) for  $P(x)=x^2$. This we shall do first.
The cases for general polynomials---which we do not need---follow by 
induction and linearity. Again we break up the argument, this time into 
5 pieces.

\begin{itemize}
\item[i)] 
We first note that $\{q,q^3\}=0$ implies via (\ref{eq:Q-Lie-preserving}) 
that $\hat q$ and $\widehat{q^3}$ commute. Since $\hat q$ and ${\hat q}^3$
commute anyway we can write $\widehat{q^3}-{\hat q}^3=A(q)$, where $A(q)$ 
takes values in the space of Hermitean operators on $\Hint$. 
\item[ii)]
We next show that $A(q)$ also commutes with $\hat p$. This follows from 
the following string of equations, where we indicated the numbers of the 
equations used in the individual steps as superscripts over the equality 
signs:
\begin{equation}
\label{eq:proof-step2-1}
[\widehat{q^3},\hat p]
\ \mathop{=}^{\sss \ref{eq:Q-Lie-preserving}}\ 
\i\hbar\widehat{\{q^3,p\}}
\ \mathop{=}^{\sss \ref{eq:PB-def}}\ 
3\i\hbar\widehat{q^2}
\ \mathop{=}^{\sss \ref{eq:squaring-law-1}}\ 
3\i\hbar{\hat q}^2
\ \mathop{=}^{\sss \ref{eq:GvH-proof-derivationrule}}\ 
[{\hat q}^3,{\hat p}]\,.
\end{equation}
Hence $A(q)$ equals a $q$-independent matrix, $\mathfrak{a}$, and we 
have 
\begin{equation}
\label{eq:eq:proof-step2-1b}
\widehat{q^3}={\hat q}^3+\mathfrak{a}\,.  
\end{equation}
\item[iii)]
We show that the matrix $\mathfrak{a}$ must actually be zero
by the following string of equations:
\begin{alignat}{1}
\widehat{q^3}
\ \mathop{=}^{\sss\ref{eq:PB-def}}\ \tfrac{1}{3}\widehat{\{q^3,qp\}}
\ \mathop{=}^{\sss \ref{eq:Q-Lie-preserving}}\ \tfrac{1}{3\i\hbar}
   [\widehat{q^3},\widehat{qp}]
\ &\mathop{=}^{\sss\ref{eq:squaring-law-3},\ref{eq:eq:proof-step2-1b}}
   \ \tfrac{1}{3\i\hbar}
   [{\hat q}^3+\mathfrak{a},\tfrac{1}{2}(\hat q\hat p+\hat p\hat q)]
   \nonumber\\ 
\label{eq:proof-step2-2}
  &\mathop{=}^{\sss\ \ *\ \ }\ 
   \tfrac{1}{6\i\hbar}[{\hat q}^3,(\hat q\hat p+\hat p\hat q)]
\  \mathop{=}^{\sss\ref{eq:GvH-proof-derivationrule}}\ {\hat q}^3\,,\qquad
\end{alignat}
where at $*$ we used that $\mathfrak{a}$ commutes with $\hat q$ and $\hat p$.
This proves (\ref{eq:gen-squaring-law-1}) for $P(q)=q^3$. Exchanging 
$p$ and $q$ and repeating the proof shows (\ref{eq:gen-squaring-law-2})
for $P(p)=p^3$. 
\item[iv)]
Using what has been just shown allows to prove 
(\ref{eq:gen-squaring-law-3}) for $P(q)=q^2$:
\begin{equation}
\label{eq:proof-step2-3}
\widehat{q^2p}
\ \mathop{=}^{\sss\ref{eq:PB-def}}\
\tfrac{1}{6}\widehat{\{q^3,p^2\}}
\ \mathop{=}^{\sss \ref{eq:Q-Lie-preserving}}\
\tfrac{1}{6\i\hbar}[\widehat{ q^3},\widehat{p^2}]
\ \mathop{=}^{\sss\ref{eq:gen-squaring-law-1},\ref{eq:squaring-law-2}}\
\tfrac{1}{6\i\hbar}[{\hat q}^3,{\hat p}^2]
\ \mathop{=}^{\sss\ref{eq:GvH-proof-derivationrule}}\
\tfrac{1}{2}({\hat q}^2\hat p+\hat p{\hat q}^2) \,.
\end{equation}
Exchanging $q$ and $p$ proves (\ref{eq:gen-squaring-law-4}) for $P(p)=p^2$. 
\item[v)]
Finally we apply the quantisation map to both sides of the classical 
equality    
\begin{equation}
\label{eq:proof-step2-4}
\tfrac{1}{9}\{q^3,p^3\}=\tfrac{1}{3}\{q^2p,p^2q\}\,.
\end{equation}
On the left hand side we replace $\widehat{q^3}$ and $\widehat{p^3}$ 
with ${\hat q}^3$ and ${\hat q}^3$ respectively and then successively 
apply (\ref{eq:GvH-proof-derivationrule}); this leads to   
\begin{equation}
\label{eq:proof-step2-5}
{\hat q}^2{\hat p}^2-2\i\hbar\hat q\hat p-\tfrac{2}{3}\hbar^2\bbbone\,.
\end{equation}
On the right hand side of (\ref{eq:proof-step2-4}) we now use 
(\ref{eq:gen-squaring-law-3}) and (\ref{eq:gen-squaring-law-4}) 
to replace $\widehat{q^2p}$ and $\widehat{p^2q}$ with 
$\tfrac{1}{2}({\hat q}^2\hat p +\hat p{\hat q}^2)$ and 
$\tfrac{1}{2}({\hat p}^2\hat q +\hat q{\hat p}^2)$ respectively 
and again successively apply (\ref{eq:GvH-proof-derivationrule}). 
This time we obtain
\begin{equation}
\label{eq:proof-step2-6}
{\hat q}^2{\hat p}^2-2\i\hbar\hat q\hat p-\tfrac{1}{3}\hbar^2\bbbone\,,
\end{equation} 
which differs from (\ref{eq:proof-step2-5}) by a term 
$-\tfrac{1}{3}\hbar^2\bbbone$. But according to (\ref{eq:Q-Lie-preserving})
both expressions should coincide, which means that we arrived at a 
contradiction. This completes part\,2 and hence the proof of the 
theorem of Groenewold and van Howe.
\end{itemize}

\subsection{Discussion}
\label{sec:discussion}
The GvH-Theorem shows that the Lie algebra of \emph{all} 
polynomials on $\reals^{2n}$ cannot be quantised (and hence no 
Lie subalgebra of $C^{\infty}(P)$ containing the polynomials). 
But its proof has also shown that the Lie \emph{sub}algebra 
\begin{equation}
\label{eq:def:funpol2}
\Funpoltwo:=\mbox{span}\left\{1,q,p,q^2,p^2,qp\right\}
\end{equation}
of polynomials of at most quadratic order \emph{can} be quantised. 
This is just the essence of the `squaring laws'  
(\ref{eq:squaring-law-1}-\ref{eq:squaring-law-3}).
 
To see that $\Funpoltwo$ is indeed a Lie subalgebra, it is sufficient 
to note that the Poisson bracket (\ref{eq:PB-def}) of a polynomial 
of $n$-th and a polynomial of $m$-th order is a polynomial of
order $(n+m-2)$. Moreover, it can be shown that $\Funpoltwo$ is a 
\emph{maximal} Lie subalgebra of $\Funpol$, i.e. that there is no 
other proper Lie subalgebra $\Fun'$ which properly contains 
$\Funpoltwo$, i.e. which satisfies $\Funpoltwo\subset\Fun'\subset\Funpol$. 

$\Funpoltwo$ contains the Lie subalgebra of all polynomials of at 
most first order:
\begin{equation}
\label{eq:def:funpol1}
\Funpolone:=\mbox{span}\left\{1,q,p\right\}\,.
\end{equation}
This is clearly a Lie ideal in $\Funpoltwo$ (not in $\Funpol$), since Poisson 
brackets between quadratic and linear polynomials are linear. $\Funpolone$ 
is also called the `Heisenberg algebra'. According to the rules 
(\ref{eq:Q-Schrodinger-consistency1}-\ref{eq:Q-Schrodinger-consistency2})
the Heisenberg algebra was required to be represented irreducibly
(cf. the discussion following (\ref{eq:Q-Schrodinger-consistency2})).
What is so special about the Heisenberg algebra? First, observe that 
it contains enough functions to coordinatise phase space, i.e. that no 
two points in phase space assign the same values to the functions 
contained in the Heisenberg algebra. Moreover, it is a minimal 
subalgebra of $\Funpol$ with this property. Hence it is a minimal 
set of classical observables whose values allow to uniquely fix 
a classical state (point in phase space). The irreducibility 
requirement can then be understood as saying that this property 
should essentially also be shared by the quantised observables,
at least up to \emph{finite} multiplicities which correspond to the 
`internal' Hilbert space $\Hint$ (a ray of which is fixed by 
finitely many eigenvalues). We will have more to say about this 
irreducibility postulate below.

The primary lesson from the GvH is that $\Funpol\subset\Funinf$
was chosen too big. It is not possible to find a quantisation map 
$\Qcal:\Funpol(+,\{{},{}\})\rightarrow \sym(\Hcal)$ which intertwines 
the Lie structures $\{{},{}\}$ and $\tfrac{1}{\i\hbar}[{},{}]$. 
This forces us to reformulate the canonical quantisation programme. 
From the discussion so far one might attempt the following rules 

\smallskip
\noindent
{\bf Rule 1.}
Given the Poisson algebra $\Funpol(+,\{{},{}\},\cdot)$ of 
all polynomials on phase space. Find a Lie subalgebra 
$\Funirr\subset\Funpol(+,\{{},{}\})$ of `basic observables' which
fulfills the two conditions: (1)~$\Funirr$ contains sufficiently 
many functions so as to coordinatise phase space, i.e. no two 
points coincide in all values of functions in $\Funirr$;
(2)~$\Funirr$ is minimal in that respect, i.e. there is 
no Lie subalgebra $\Funirr'$ properly contained in $\Funirr$ which also 
fulfills~(1). 

\smallskip
\noindent
{\bf Rule 2.}
Find another Lie subalgebra $\Funquant\subset\Funpol(+,\{{},{}\})$ so 
that $\Funirr\subseteq\Funquant$ and that $\Funquant$ can be quantised, 
i.e. a Lie homomorphism $\Qcal:\Funquant\rightarrow\sym(\Hcal)$
can be found, which intertwines the Lie structures $\{{},{}\}$ and 
$\frac{1}{\i\hbar}[{},{}]$. Require $\Qcal$ to be such that 
$\Qcal(\Funirr)$ act almost irreducibly, i.e. up to finite multiplicity, 
on $\Hcal$. Finally, require that $\Funquant$ be maximal in $\Funpol$, 
i.e. that there is no $\Funquant'$ with 
$\Funquant\subset\Funquant'\subset\Funpol(+,\{{},{}\})$. 

Note that the choice of $\Funquant$ is generally far from unique. 
For example, instead of choosing $\Funquant=\Funpoltwo$, i.e. the 
polynomials of at most quadratic order, we could choose 
$\Funquant=\Funpolinfone$, the polynomials of at most linear order 
in momenta with coefficients which are arbitrary polynomials in $q$. 
A general element in $\Funpolinfone$ has the form 
\begin{equation}
\label{eq:funinfone-1}
f(q,p)=g(q)+h(q)\,p  
\end{equation}
where $g,h$ are arbitrary polynomials with real coefficients. 
The Poisson bracket of two such functions is 
\begin{equation}
\label{eq:funinfone-2}
\{f_1,f_2\}=\{g_1+h_1p,g_2+h_2p\}=g_3+h_3p\,,
\end{equation}
where
\begin{equation}
\label{eq:funinfone-3}
g_3=g'_1h_2-g'_2h_1\quad\mbox{and}\quad h_3=h'_1h_2-h_1h'_2\,.
\end{equation}
The quantisation map applied to $f$ is then given by 
\begin{equation}
\label{eq:funinfone-4}
\widehat{f}=g(\hat q)-\i\hbar(\tfrac{1}{2}h'(\hat q)
+h(\hat q)\ \tfrac{d}{dq})\,,
\end{equation}
where $h'$ denotes the derivative of $h$ and $\hat q$ and $\hat p$ 
are just the Schr\"odinger operators `multiplication by $q$' and 
`$-\i\hbar d/dq$' respectively. The derivative term proportional to 
$h'$ is necessary to make $\widehat{f}$ symmetric (an overline 
denoting complex conjugation):
\begin{equation}
 \begin{split}
 \label{eq:funinfone-5}
 [\tfrac{\i}{2}h'(q)\psi(q)+\i h(q){\psi'}(q)]\,\overline{\phi(q)}
\,&=\,
 \psi(q)\,[\overline{\tfrac{\i}{2}h'(q)\phi(q)+\i h(q)\phi'(q)}]\\
&+\,(\i h\psi\overline\phi)'(q)\,,\\
\end{split}
\end{equation}
where the last term vanishes upon integration. Moreover, a simple 
computation readily shows that the map $f\mapsto\hat f$ indeed defines 
a Lie homomorphism from $\Funpolinfone$ to $\sym(\Hcal)$:
\begin{equation}
\label{eq:funinfone-6}
\tfrac{1}{\i\hbar}[\hat f_1,\hat f_2]=
g_3(q)-\i\hbar\left(\tfrac{1}{2}h'_3(q)+h_3(q)\tfrac{d}{dq}\right)
=\widehat{\{f_1,f_2\}}\,,
\end{equation}
with $f_{1,2}$ and $g_{3},h_3$ as in (\ref{eq:funinfone-2}) and 
(\ref{eq:funinfone-3}) respectively. Hence (\ref{eq:funinfone-4})
gives a quantisation of $\Funpolinfone$.  

It can be shown (\cite{Gotay:1999}, Thm.\,8) that $\Funpoltwo$ and 
$\Funpolinfone$ are the only maximal Lie subalgebras of $\Funpol$ 
which contain the Heisenberg algebra $\Funpolone$. In this sense, 
if one restricts to polynomial functions, there are precisely two 
inextendible quantisations. 

So far we restricted attention to polynomial functions. Since $\Funpol$
is already too big to be quantised, there is clearly no hope to quantise 
all $C^{\infty}$ functions on our phase space $\reals^{2n}$. 
For general phase spaces $P$ (i.e. not isomorphic to $\reals^{2n}$)  
there is generally no notion of `polynomials' and hence no simple way to 
characterise suitable Lie subalgebras of $\Funinf(+,\{{},{}\})$. 
But experience with the GvH Theorem suggests anyway to conjecture that, 
subject to some irreducibility postulate for some minimal choice of 
$\Funirr\subset\Funinf$, there is \emph{never} a quantisation of 
$\Funinf$. (A quantisation of all $C^{\infty}$ functions is called 
\emph{full quantisation} in the literature.) Surprisingly there is a 
non-trivial counterexample to this conjecture: it has been shown that 
a full quantisation exists for the 2-torus \cite{Gotay:1996-2torus}. 
One might first guess that this is somehow due to the compactness of 
the phase space. But this is not true, as a GvH obstruction to full 
quantisation does exist for the 2-sphere \cite{Gotay:1996-2sphere}. 
But the case of the 2-torus seems exceptional, even mathematically. 
The general expectation is indeed that GvH-like obstructions are in 
some sense generic, though, to my knowledge, there is no generally 
valid formulation and corresponding theorem to that effect. 
(For an interesting early attempt in this direction see \cite{Gotay-1980}.) 
Hence we face the problem to determine $\Funirr$ and $\Funquant$ within 
$\Funinf$. There is no general theory how to do this. If $P$ is 
homogeneous, i.e. if there is a finite dimensional Lie group $G$
(called the `canonical group') that acts transitively on $P$ and 
preserves the Poisson bracket (like the $2n$ translations in 
$\reals^{2n}$) one may generate $\Funirr$ from the corresponding 
momentum maps. This leads to a beautiful theory 
\cite{Isham-LesHouches:1984} for such homogeneous situations, but 
general finite dimensional $P$ do not admit a finite dimensional 
canonical group $G$, and then things become much more complicated.

\subsection{The r\^ole of the irreducibility-postulate}
\label{sec:irreducibility}
\begin{definition} Quantisation without the irreducibility 
postulate 
(\ref{eq:Q-Schrodinger-consistency1}-\ref{eq:Q-Schrodinger-consistency2})
is called \emph{pre-quantisation}.
\end{definition}
Given the GvH result, the following is remarkable:
\begin{theorem}
A prequantisation of the Lie algebra $\Funinf(+,\{{},{}\})$ of 
all $C^{\infty}$-functions on $\reals^{2n}$ exists. 
\end{theorem}
\noindent
The proof is constructive by means of \emph{geometric quantisation}. 
Let us briefly recall the essentials of this approach: The Hilbert 
space of states is taken to be $\Hcal=L^2(\reals^{2n},\,d^nqd^np)$,
i.e. the square integrable functions on \emph{phase} space 
($2n$ coordinates), instead of configuration space ($n$ coordinates). 
The quantisation map is as follows\footnote{
% FOOTENOTETEXT
Unlike in ordinary Schr\"odinger quantisation, where 
$\vert\psi(q)\vert^2$ is the probability density for the system in 
configuration space, the corresponding quantity 
$\vert\psi(q,p)\vert^2$ in geometric quantisation has \emph{not} the 
interpretation of a probability density in phase space. The formal 
reason being that in geometric quantisation $\hat q$ is not just a 
multiplication operator (cf. (\ref{eq:geo-quant-6})). For example, 
if $\psi$ has support in an arbitrary small neighbourhood $U$ of 
phase space this does not mean that we can simultaneously reduce the 
uncertainties of $\hat q$ and $\hat p$, since this would violate the 
uncertainty relations which hold unaltered in geometric quantisation. 
Recall that the uncertainty relations just depend on the commutation 
relations since they derive from the following generally valid formula 
by dropping the last term:
($\langle\cdot\rangle_{\psi}$ denotes the expectation value in the 
state $\psi$, $[\cdot,\cdot]_+$ the anticommutator and 
$\hat f_{\sss 0}:=\hat f-\langle\hat f\rangle_\psi\bbbone$):  
\begin{equation}
\label{eq:uncertainty}
\langle{\hat f}_{\sss 0}^2\rangle_{\psi}
\langle{\hat g}_{\sss 0}^2\rangle_{\psi}\geq
\tfrac{1}{4}\left\{
\vert\langle[\hat f,  \hat g]    \rangle_\psi\vert^2 + 
\vert\langle[\hat f_{\sss 0},\hat g_{\sss 0}]_+\rangle_\psi\vert^2
\right\}\,.
\end{equation}
}:
\begin{equation}
\label{eq:geo-quant-1}
\Qcal(f)=\i\hbar\nabla_{X_f}+f\,,
\end{equation}
where $\nabla$ is a `covariant-derivative' operator, which is 
\begin{equation}
\label{eq:geo-quant-2}
\nabla=d+A\,.
\end{equation}
Here $d$ is just the ordinary (exterior) derivative and the connection
1-form, $A$, is proportional to the canonical 1-form 
(cf. (\ref{eq:symp-6})) $\theta:=p_i\,dq^i$: 
\begin{equation}
\label{eq:geo-quant-3} 
A=-\tfrac{\i}{\hbar}\,\theta=-\tfrac{\i}{\hbar}p_i\,dq^i\,.
\end{equation}
The curvature, $F=dA$, is then proportional to the symplectic 2-form
$\omega=d\theta$:
\begin{equation}
\label{eq:geo-quant-4}
F=\tfrac{\i}{\hbar}\omega=\tfrac{\i}{\hbar}dq^i\wedge dp_i\,.
\end{equation}
If $X_f$ is the Hamiltonian vector field on phase space associated 
to the phase-space function $f$ (cf. (\ref{eq:symp-1})), then in 
canonical coordinates it has the form 
\begin{equation}
\label{eq:geo-quant-5} 
X_f=(\partial_{p_i}f)\partial_{q^i}-(\partial_{q^i}f)\partial_{p_i}\,.
\end{equation}
The map $f\mapsto X_f$ is a Lie homomorphism from 
$\Funinf(+,\{{},{}\})$ to the Lie algebra of vector fields on phase 
space, i.e. $X_{\{f,g\}}=[X_f,X_g]$. The operator $\hat f$ is formally 
self-adjoint and well defined on Schwarz-space (rapidly decreasing 
functions), which we take as our invariant dense domain $\Dcal$. 
Explicitly its action reads:
\begin{equation}
\label{eq:geo-quant-5b}
\hat f=i\hbar
\bigl((\partial_{q^i}f)\partial_{p_i}-
     (\partial_{p_i}f)\partial_{q^i}\bigr)+
\bigl(f-(\partial_{p_i}f)p_i\bigr)\,,
\end{equation}
which clearly shows that all operators are differential operators of at 
most degree one. This makes it obvious that a squaring-law in the form 
$\hat f\hat g=\widehat{fg}$ \emph{never} applies. For example, for $n=1$ 
we have for $\hat q,\hat p$ and their squares:
\begin{alignat}{3}
\label{eq:geo-quant-6}
\hat q\,&=\,q+\i\hbar\partial_p\,,\quad 
&&\widehat{q^2}=&&q^2+2\i\hbar\,\partial_p\,,\\
\hat p\,&=-\i\hbar\partial_q\,,   \quad 
&&\widehat{p^2}=-&&p^2-2\i\hbar\,p\partial_q\,.
\end{alignat}
One now proves by direct computation that (\ref{eq:geo-quant-1})
indeed defines a Lie homomorphism:
\begin{equation}
\begin{split}
\label{eq:geo-quant-7}
\tfrac{1}{\i\hbar}\,[\Qcal(f),\Qcal(g)]
&\,=\,\tfrac{1}{\i\hbar}\,[\i\hbar\nabla_{X_f}+f\,,\,\i\hbar\nabla_{X_g}+g]\\
&\,=\, \i\hbar\,[\nabla_{X_f},\nabla_{X_g}]+X_f(g)-X_g(f)\\
&\,=\,\i\hbar\,\bigl(\nabla_{[X_f,X_g]}+F(X_f,X_g)\bigr)+2\{f,g\}\\
&\,=\,\i\hbar\, \nabla_{X_{\{f,g\}}}+\{f,g\}=\Qcal(\{f,g\})\,,\\
\end{split}
\end{equation}
where we just applied the standard identity for the curvature of the  
covariant derivative (\ref{eq:geo-quant-2}):  
$F(X,Y)=\nabla_X\nabla_Y-\nabla_Y\nabla_X-\nabla_{[X,Y]}$ and also used  
$-\i\hbar\,F(X_f,X_g)=\omega(X_f,X_g)=\{f,g\}$
(cf. (\ref{eq:symp-4})). 

Let us now look at a simple specific example: the linear harmonic 
oscillator. We use units where its mass and angular frequency 
equal~1. The Hamiltonian function and vector field are then given by:
\begin{equation}
\label{eq:har-osz-1}
H=\tfrac{1}{2}(p^2+q^2)\,\Rightarrow\, X_H=p\partial_q-q\partial_p\,,
\end{equation}
whose quantisation according to (\ref{eq:geo-quant-1}) is
\begin{equation}
\label{eq:har-osz-2}
\hat H=-\i\hbar\left(p\partial_q-q\partial_p\right)+
\tfrac{1}{2}\left(q^2-p^2\right)\,.
\end{equation}
Introducing polar coordinates on phase space: $q=r\cos(\varphi)$
$p=r\sin(\varphi)$, the Hamiltonian becomes 
\begin{equation}
\label{eq:har-osz-3}
\hat H=\i\hbar\partial_{\varphi}+\tfrac{r^2}{2}\cos(2\varphi)\,.
\end{equation}      
The eigenvalue equation reads
\begin{equation}
\label{eq:har-osz-4}
\hat H\psi=E\psi\ \Leftrightarrow\ 
\partial_{\varphi}\psi=
-\tfrac{\i}{\hbar}\left(E-\tfrac{r^2}{2}\cos(2\varphi)\right)\,\psi\,,
\end{equation}
whose solution is 
\begin{equation}
\label{eq:har-osz-5}
\psi(r,\varphi)=\psi_0(r)\,
\exp\left\{
-\tfrac{\i}{\hbar}\left(E\varphi-\tfrac{r^2}{2}\,\sin(2\varphi)\right)
\right\}\,,
\end{equation}
where $\psi_0$ is an arbitrary function in $L^2(\reals_+,rdr)$.
Single valuedness requires 
\begin{equation}
\label{eq:har-osz-6} 
E=E_n=n\hbar\,,\quad n\in\integers\,,
\end{equation}
with each energy eigenspace being isomorphic to the space of 
square-integrable functions on the positive real line with respect 
to the measure $rdr$:
\begin{equation}
\label{eq:har-osz-7}
\Hcal_n= L^2(\reals_+,rdr)\,.
\end{equation}

Hence we see that the difference to the usual Schr\"odinger 
quantisation is not simply an expected degeneracy of the energy 
eigenspaces which, by the way, turns out to be quite enormous, 
i.e. infinite dimensional for each energy level. What is much 
worse and perhaps less expected is the fact that the energy spectrum 
in prequantisation is a proper extension of that given by Schr\"odinger
quantisation and, in distinction to the latter, that it is 
\emph{unbounded from below}. This means that there is no ground state 
for the harmonic oscillator in prequantisation which definitely 
appears physically wrong. Hence there seems to be some deeper physical 
significance to the irreducibility postulate than just mere avoidance 
of degeneracies.

%%%%%%%%%%%%%%%%%%%%%%%% SECTION 2: CONSTRAINED SYSTEMS %%%%%%%%%%%%%%%%%%%%

\section{Constrained Systems}
For systems with gauge redundancies\footnote{We deliberately avoid the 
word `symmetry' in this context, since the action of a gauge group has 
a completely different physical interpretation than the action of a 
proper symmetry; only the latter transforms states into other, physically 
distinguishable states. See section\,6.3 in \cite{Joos-etal:2003} 
for a more comprehensive discussion of this point.} the original phase space 
$P$ does not directly correspond to the set of (mutually different) 
classical states. First of all, only a subset $\hat P\subset P$ will 
correspond to classical states of the system, i.e. the system is 
\emph{constrained} to $\hat P$. Secondly,
the points of $\hat P$ label the states of the systems in a redundant 
fashion, that is, one state of the classical system is labeled by many 
points in $\hat P$. The set of points which label the same state form
an orbit of the group of gauge transformations which acts on $\hat P$. 
`Lying in the same orbit' defines an equivalence relation (denoted by $\sim$)
on $\hat P$ whose equivalence classes form the space 
$\bar P:=\hat P/\!\!\sim$ which is called the \emph{reduced phase space}. 
Its points now label the classical states in a faithful fashion. Note 
that it is a quotient-space of the sub-space $\hat P$ of $P$ and can, in 
general, therefore not be represented as a subspace of $P$.    

A straightforward strategy to quantise such a system is to `solve' the 
constraints, that is, to construct $\bar P$. One could then apply the 
same methods as for unconstrained systems, at least as long as 
$\bar P$ will be a $C^{\infty}$-manifold with a symplectic structure 
(cf. Appendix\,1).\footnote{
In passing we remark that even though $P$ may (and generally is in 
applications) a cotangent bundle $T^*Q$ for some configuration space 
$Q$, this need not be true for $\bar P$ , i.e. there will be no space 
$\bar Q$ such that $\bar P\cong T^*\bar Q$. For this reason it is 
important to develop quantisations strategies that apply to general 
symplectic manifolds.} In particular, we can then consider the Poisson
algebra of $C^{\infty}$-functions and proceed as for unconstrained 
systems. 

However, in general it is analytically very difficult to explicitly 
do the quotient construction $\hat P\rightarrow \hat P/\!\!\sim\ =\bar P$,
i.e. to solve the constraints \emph{classically}. Dirac has outlined a 
strategy to implement the constraints \emph{after} 
quantisation~\cite{Dirac:1964}. The basic mathematical reason why this 
is considered a simplification is seen in the fact that the whole problem 
is now posed in \emph{linear} spaces, i.e. the construction of sub- 
and quotient spaces in the (linear) spaces of states and observables. 

Dirac's ideas have been reviewed, refined, and discussed many times 
in the literature; see e.g. the comprehensive textbook by Henneaux and 
Teitelboim~\cite{Henneaux-Teitelboim:1992}. Here we shall merely give 
a brief coordinate-free description of how to construct the right 
classical Poisson algebra of functions (the `physical observables').

\subsection{First-class constraints}
Let $(P,\omega)$ be a symplectic manifold which is to be thought of as 
an initial phase space of some gauge system. The physical states then
correspond to the points of some submanifold $\hat P\hookrightarrow P$.
Usually $\hat P$ is characterised as zero-level set of some given 
collection of functions, $\hat P=\{p\in P\mid\phi_\alpha(p)=0,\,
\alpha =1,2,...,\mbox{codim}(\hat P)\}$, where 
$\mbox{codim}(\hat P):=\mbox{dim}(P)-\mbox{dim}(\hat P)$ 
denotes the `codimension' of $\hat P$ in $P$. The ensuing 
formulae will then depend on the choice of $\phi_\alpha$, though the 
resulting theory should only depend on $\hat P$ and not on its analytical 
characterisation. To make this point manifest we just work with the 
geometric data. As usual, we shall denote the tangent bundles of $P$ 
and $\hat P$ by $TP$ and $T\hat P$ respectively. The restriction of $TP$ 
to $\hat P$ (which also contains vectors not tangent to $\hat P$) is 
given by $TP\vert_{\sss\hat P}:=\{X\in T_pP\mid p\in \hat P\}$. 
The $\omega$-orthogonal complement of $T_p\hat P$ is now defined as
follows: 
\begin{equation}
\label{eq:const-1}
T_p^{\perp}\hat P:=\{X\in T_pP\vert_{\sss\hat P}\mid 
                     \omega(X,Y)=0,\,\forall Y\in T_pP\}\,.
\end{equation}
\begin{definition}
A submanifold $\hat P\hookrightarrow P$ is called \emph{coisotropic} iff 
$T^{\perp}\hat P\subset T\hat P$.
\end{definition}
Since $\omega$ is non degenerate we have 
$\dim T_p\hat P+\dim T_p^{\perp}\hat P=\dim T_pP$, hence 
$\dim T_p^{\perp}\hat P=\mbox{codim}\,\hat P$. This means that 
for coisotropic embeddings $i:\hat P\hookrightarrow P$ the 
kernel\footnote{The kernel (or `null-space') of a bilinear form 
$f$ on $V$ is the subspace 
$\kernel(f):=\{X\in V\mid f(X,Y)=0,\ \forall\, Y\in V\}$.}
of the pulled-back symplectic form $\hat\omega:=i^*\omega$ on $\hat P$ 
has the maximal possible number of dimensions, namely $\mbox{codim}\,\hat P$.
\begin{definition}
A constrained system $\hat P\hookrightarrow P$ is said to be of \emph{first 
class} iff $\hat P$ is a coisotropic submanifold of $(P,\omega)$.
\end{definition}
From now on we consider only first class constraints.

\begin{lemma}
\label{lemma:const-1}
$T^{\perp}\hat P\subset TP\vert_{\sss\hat P}$ is an integrable subbundle. 
\end{lemma}
\begin{proof}
The statement is equivalent to saying that the commutator of any two 
$T^{\perp}\hat P$-valued vector fields $X,Y$ on $\hat P$ is again 
$T^{\perp}\hat P$-valued. Using $[X,Y]=L_XY$ and formula (\ref{eq:symp-3})
we have\footnote{\label{fnote:insert-conv} We shall use the symbol 
$\Insert$ to denote the insertion of a vector (standing to the left 
of $\Insert$) into the first slot of a form (standing to the right 
of $\Insert$). For example, for the 2-form $\omega$, $X\Insert\omega$ 
denotes the 1-form $\omega(X,\cdot)$.} 
$[X,Y]\Insert\hat\omega= L_X(Y\Insert\hat\omega)
-Y\Insert L_X\hat\omega=-Y\Insert d(X\Insert\hat\omega)=0$, 
since $Y\Insert\hat\omega=0=X\Insert\hat\omega$ and 
$d\hat\omega=di^*\omega=i^*d\omega=0$ due to $d\omega=0$. 
\end{proof}

\begin{definition}
The \emph{gauge algebra}, $\Gau$, is defined to be the set of all 
functions (out of some function class $\Fun$, usually $C^{\infty}(P)$)
which vanish on $\hat P$:
\begin{equation}
\label{eq:const-2}
\Gau:=\{f\in\Fun(P)\mid f\vert_{\sss\hat P}\equiv 0\}\,.
\end{equation}
\end{definition}
$\Gau$ uniquely characterises the constraint surface $\hat P$ in a 
coordinate independent fashion. In turn, this allows to characterise 
the constraints algebraically; $\Gau$ is in fact a Poisson 
algebra. To see this, first note that it is obviously an ideal of
the associative algebra $\Fun(+,\cdot)$, since any pointwise product with 
an element in $\Gau$ also vanishes on $\hat P$. Next we show 
\begin{lemma}
\label{lemma:const-2}
$f\in\Gau$ implies that $X_f\vert_{\sss\hat P}$ is $T^{\perp}\hat P$-valued.
\end{lemma}
\begin{proof}
$f\vert_{\sss\hat P}\equiv 0
\Rightarrow\kernel(df\vert_{\sss\hat P})
=\kernel((X_f\Insert\omega)\vert_{\sss\hat P})\supseteq T\hat P$. 
Hence $X_f\vert_{\sss\hat P}$ is $T^{\perp}\hat P$-valued.
\end{proof}

Now it is easy to see that $\Gau$ is also a Lie algebra, since 
for $f,g\in\Gau$ we have 
\begin{equation}
\label{eq:const-3}
\{f,g\}\vert_{\sss\hat P}
=X_f(g)\vert_{\sss\hat P}
=X_f\Insert dg\vert_{\sss\hat P}
=X_g\Insert X_f\Insert\omega\vert_{\sss\hat P}=0\,,
\end{equation}
where (\ref{eq:symp-1}) and Lemma\,\ref{lemma:const-2} was used 
in the last step.
Hence $\Gau$ is shown to be an associative and Lie algebra, hence 
a Poisson algebra. But note that whereas it is an associative ideal 
it is not a Lie ideal. Indeed, for $f\in\Gau$ and $g\in\Fun$ we 
have $\{f,g\}\vert_{\sss\hat P}=X_f(g)\vert_{\sss\hat P}\not =0$ 
for those $g$ which vary on $\hat P$ in the direction of $X_f$. 

The interpretation of $\Gau$ is that its Hamiltonian vector fields 
generate gauge transformations, that is, motions which do not correspond
to physically existing degrees of freedom. Two points in $\hat P$ which 
are on the same connected leaf of $T^{\perp}\hat P$ correspond to the 
\emph{same} physical state. The observables for the system described by 
$\hat P$ must therefore Poisson-commute with all functions in $\Gau$. 
Hence one might expect the Poisson algebra of physical observables 
to be given by the quotient $\Fun/\Gau$. However, since $\Gau$ is not 
a Lie ideal in $\Fun$ the quotient is not a Lie algebra and hence
not a Poisson algebra either.  The way to proceed is to consider the 
biggest Poisson subalgebra of $\Fun$ which contains $\Gau$ as Lie ideal 
and then take the quotient. Hence we make the following 
\begin{definition}
The \emph{Lie idealiser} of $\Gau\subset\Fun$ is 
\begin{equation}
\label{eq:const-4}
\Ical_{\sss\Gau}:=\{f\in\Fun\mid\{f,g\}\vert_{\sss\hat P}=0,
\ \forall g\in\Gau\}\,.
\end{equation}      
\end{definition}
$\Ical_{\sss\Gau}$ is the space of functions which, in Dirac's terminology
\cite{Dirac:1964}, are said to \emph{weakly} commute with all gauge 
functions $g\in\Gau$; that is, $\{f,g\}$ is required to vanish only
\emph{after} restriction to $\hat P$. 

\begin{lemma}
$\Ical_{\sss\Gau}$ is a Poisson subalgebra of $\Fun$ which contains 
$\Gau$ as ideal.
\end{lemma}
\begin{proof}
Let $f,g\in\Ical_{\sss\Gau}$ and $h\in\Gau$. Then clearly 
$f+g\in\Ical_{\sss\Gau}$ and also 
$\{f\cdot g,h\}\vert_{\sss\hat P}=f\cdot\{g,h\}\vert_{\sss\hat P}
 +g\cdot\{f,h\}\vert_{\sss\hat P}=0$ (since each term vanishes), 
hence $\Ical_{\sss\Gau}$ is an associative subalgebra. Moreover,
using the Jacobi identity, we have 
\begin{equation}
\{\{f,g\},h\}\vert_{\sss\hat P}
=\{\underbrace{\{h,g\}}_{\in\Gau},f\}\vert_{\sss\hat P}
+\{\underbrace{\{f,h\}}_{\in\Gau},g\}\vert_{\sss\hat P}=0\,,
\end{equation}
which establishes that $\Ical_{\sss\Gau}$ is also a Lie subalgebra. 
$\Gau$ is obviously an associative ideal in $\Ical_{\sss\Gau}$ 
(since it is such an ideal in $\Fun$) and, by definition, also a Lie 
ideal. Hence it is a Poisson ideal. 
\end{proof}

It follows from its very definition that $\Ical_{\sss\Gau}$ is maximal 
in the sense that there is no strictly larger subalgebra in $\Fun$ in 
which $\Gau$ is a Poisson algebra. Now we can define the algebra
of \emph{physical observables}:
\begin{definition}
The \emph{Poisson algebra of physical observables} is given by 
\begin{equation}
\label{eq:const-6}
\Ophys:=\Ical_{\sss\Gau}/\Gau\,.
\end{equation}
\end{definition} 

Since the restriction to $\hat P$ of a Hamiltonian vector field $X_g$ 
is tangent to $\hat P$ if $g\in\Gau$ (by Lemma\,\ref{lemma:const-2} and 
coisotropy), we have  
\begin{equation}
\label{eq:const-7}
\begin{split}
\Ical_{\sss\Gau}
=&\{f\in\Fun\mid X_g(f)\vert_{\sss\hat P}=0,\forall g\in\Gau\}\\ 
=&\{f\in\Fun\mid X_g\vert_{\sss\hat P}(f\vert_{\sss\hat P})=0,
\forall g\in\Gau\}\,,\\
\end{split}
\end{equation}
which shows that $\Ical_{\sss\Gau}$ is the subspace of all functions 
in $\Fun$ whose restrictions to $\hat P$ are constant on each connected 
leaf of the foliation tangent to the integrable subbundle 
$T^{\perp}\hat P$. If the space of leaves is a smooth manifold\footnote{
The `space of leaves' is the quotient space with respect to the equivalence 
relation `lying on the same leaf'. If the leaves are the orbits of a group 
action (the group of gauge transformations) then this quotient will be a 
smooth manifold if the group action is smooth, proper, and free
(cf.~ Sect.\,4.1 of \cite{Abraham-Marsden:1978}).}  
it has a natural symplectic structure. In this case it is called the 
\emph{reduced phase space} $(\bar P,\bar\omega)$. $\Ophys$ can then be 
naturally identified with the Poisson algebra of (say $C^{\infty}$-) 
functions thereon.   

We finally mention that instead of the Lie idealiser $\Ical_{\sss\Gau}$
we could not have taken the Lie centraliser
\begin{equation}
\label{eq:const-8}
\begin{split}
\Ccal_{\sss\Gau}:=&\{f\in\Fun\mid\{f,g\}=0,\ \forall g\in\Gau\}\\
                 =&\{f\in\Fun\mid X_g(f)=0,\ \forall g\in\Gau\}\,, \\
\end{split}
\end{equation}
which corresponds to the space of functions which, in Dirac's 
terminology~\cite{Dirac:1964}, \emph{strongly} commute with all 
gauge functions. This space is generally far too small, as can be 
seen from the following 
\begin{lemma}
If $\hat P$ is a closed subset of $P$ we have
\begin{equation}
\label{eq:const-9}
\mbox{Span}\{X_g(p),\ g\in\Gau\}=
\begin{cases}
T_p^{\perp}\hat P &\mbox{for}\ p\in\hat P \\
T_pP              &\mbox{for}\ p\in P-\hat P\,.
\end{cases}
\end{equation}
\end{lemma}
\begin{proof}
For $p\in\hat P$ we know from Lemma\,\ref{lemma:const-2} that 
$X_g(p)\in T_p^{\perp}\hat P$. Locally we can always find 
$\mbox{codim}(\hat P)$ functions $g_i\in \Gau$ whose differentials 
$dg_i$ (and hence whose vector fields $X_{g_i}$) at $p$ are 
linearly independent. To see that the $X_g(p)$ span all of $T_pP$ for 
$p\not\in\hat P$, we choose a neighbourhood $U$ of $p$ such that 
$U\cap\hat P=\emptyset$ (such $U$ exists since $\hat P\subset P$ is 
closed by hypothesis) and $\beta\in C^{\infty}(P)$ such that 
$\beta\vert_{\sss U}\equiv 1$ and $\beta\vert_{\sss \hat P}\equiv 0$.
Then $\beta\cdot h\in\Gau$ for all $h\in C^{\infty}(P)$ and  
$(\beta\cdot h)\vert_{\sss U}=h\vert_{\sss U}$, which shows that  
$\mbox{Span}\{X_g(p), g\in\Gau\}=\mbox{Span}\{X_g(p), g\in C^{\infty}(P)\}
=T_pP$.
\end{proof}
This Lemma immediately implies that functions which strongly commute 
with all gauge functions must have altogether vanishing directional 
derivatives outside $\hat P$, that is, they must be constant on any 
connected set outside $\hat P$. By continuity they must be also 
constant on any connected subset of $\hat P$. Hence the condition of 
strong commutativity is far too restrictive. 

Sometimes strong commutativity is required, but only with a somehow 
preferred subset $\phi_\alpha$, $\alpha=1,\cdots,\mbox{codim}(\hat P)$, 
of functions in $\Gau$; for example, the component functions of a 
momentum map (cf. Sect.\,4.2 of \cite{Abraham-Marsden:1978}) of a group 
(the group of gauge transformations) that acts symplectomorphically 
(i.e. $\omega$-preserving) on $P$. The size of the space of functions 
on $P$ that strongly commute with all $\phi_\alpha$ will generally depend
delicately on the behaviour of the $\phi_\alpha$ off the constraint 
surface, and may again turn out to be too small. The point being that 
even though the leaves generated by the $\phi_\alpha$ may behave well 
\emph{within} the zero-level set of all $\phi_\alpha$ (the constraint 
surface), so that sufficiently many invariant (i.e. constant along the 
leaves) functions exist on the constraint surface, the leaves may become 
more `wild' in infinitesimal neighbouring level sets, thereby 
forbidding most of these functions to be extended to some invariant 
functions in a neighbourhood of $\hat P$ in $P$. See Sect.\,3 of 
\cite{Bordemann-etal:2000} for an example and more discussion of this 
point.

%%%%%%%%%%%%%%%%%%%%%%%% APPENDICES %%%%%%%%%%%%%%%%%%%%%%%%%%%%%%%%%%%%%%%%
\appendix

%%%%%%%% APPENDIX 1

\section*{Appendix 1: Geometry of Hamiltonian Systems}  
\label{sec:appendix-2}
A \emph{symplectic manifold} is a pair $(P,\omega)$, where $P$ is a 
differentiable manifold and $\omega$ is a closed (i.e. $d\omega=0$) 
2-form which is non-degenerate 
(i.e. $\omega_p(X_p,Y_p)=0,\,\forall X_p\in T_pP$, implies $Y_p=0$
for all $p\in P$). The last condition implies that $P$ is even 
dimensional. Let $C^{\infty}(P)$ denote the set of infinitely
differentiable, real valued functions on $P$ and $\Xcal(P)$ the 
set of infinitely differentiable vector fields on $P$. $\Xcal(P)$ is a 
real Lie algebra (cf. Appendix\,2) whose Lie product is the 
commutator of vector fields. There is a map 
$X:C^{\infty}(P)\rightarrow \mathcal{X}(P)$, $f\mapsto X_f$, uniquely 
defined by\footnote{For notation recall footnote\,\ref{fnote:insert-conv}.}  
\begin{equation}
\label{eq:symp-1}
X_f\Insert\omega=-df \,.  
\end{equation}
The kernel of $X$ in $C^{\infty}(P)$ are the constant functions and the 
image of $X$ in $\Xcal(P)$ are called Hamiltonian vector fields.
The Lie derivative of $\omega$ with respect to an Hamiltonian 
vector field is always zero: 
\begin{equation}
\label{eq:symp-2}
L_{X_f}\omega=d(X_f\Insert\omega)=-ddf=0\,,
\end{equation}
where we used the following identity for the Lie derivative $L_Z$ with 
respect to any vector field $Z$ on forms of any degree: 
\begin{equation}
\label{eq:symp-3}
L_Z=d\circ (Z\Insert) +\, (Z\Insert)\circ d\,.
\end{equation}
The map $X$ can be used to turn $C^{\infty}$ into a Lie algebra.  
The Lie product $\{\cdot,\cdot\}$ on $C^{\infty}$ is called Poisson 
bracket and defined by 
\begin{equation}
\label{eq:symp-4}  
\{f,g\}:=\omega(X_f,X_g)=X_f(g)=-X_g(f)\,,
\end{equation}
where the 2nd and 3rd equality follows from (\ref{eq:symp-1}). With 
respect to this structure the map $X$ is a homomorphism of Lie 
algebras:
\begin{equation}
\begin{array}{rcl}
\label{eq:symp-5}
X_{\{f,g\}}\Insert\omega=-d\{f,g\}
 &\displaystyle{\mathop{=}^{\sss\ref{eq:symp-4}}}
 & -d(X_g\Insert X_f\Insert\omega)\\ 
 &\displaystyle{\mathop{=}^{\sss\ref{eq:symp-3},\ref{eq:symp-1}}}
 & - L_{X_g}(X_f\Insert\omega)\\
 &\displaystyle{\mathop{=}^{\sss\ref{eq:symp-2}}}
 &[X_f,X_g]\Insert\omega\,.
\end{array}
\end{equation}
One may say that the map $X$ has pulled back the Lie structure 
from $\Xcal(P)$ to $C^{\infty}(P)$. Note that (\ref{eq:symp-5})
also expresses the fact that Hamiltonian vector fields form 
a Lie subalgebra of $\Xcal(P)$

Special symplectic manifolds are the cotangent bundles. Let $M$
be a mani\-fold and $P=T^*M$ its cotangent bundle with projection 
$\pi:T^*M\rightarrow M$. On $P$ there exists a naturally given 
1-form field (i.e. section of $T^*P=T^*T^*M)$), called the 
canonical 1-form (field) $\theta$:
\begin{equation}
\label{eq:symp-6}
\theta_p:=p\circ\pi_{*}\vert_p\,.
\end{equation}
In words, application of $\theta$ to $Z_p\in T_pP$ is as
follows: project $Z_p$ by the differential $\pi_*$, evaluated at $p$,
into $T_xM$, where $x=\pi(p)$, and then act upon it by $p$, 
where $p\in\pi^{-1}(x)=T^*_xM$ is understood as 1-form on $M$. 
The exterior differential of the canonical 1-form defines a symplectic 
structure on $P$ (the minus sign being conventional):
\begin{equation}
\label{eq:symp-7}
\omega:=-d\theta\,.
\end{equation}

In canonical (Darboux-) coordinates 
($\{q^i\}$ on $M$ and $\{p_i\}$ on the fibres $\pi^{-1}(x)$)
one has 
\begin{equation}
\label{eq:symp-8}
\theta=p_i\,dq^i\quad\mbox{and}\quad\omega=dq^i\wedge dp_i\,,
\end{equation}
so that 
\begin{equation}
\label{eq:symp-9}
\{f,g\}=\sum_i
\left(
 \frac{\partial f}{\partial q^i}\frac{\partial g}{\partial p_i}
-\frac{\partial f}{\partial p_i}\frac{\partial g}{\partial q^i}
\right)\,.
\end{equation}
In this coordinates the Hamiltonian vector field $X_f$ reads:
\begin{equation}
\label{eq:symp-10}
X_f=(\partial_{p_i}f)\partial_{q^i}-
    (\partial_{q^i}f)\partial_{p_i}\,.
\end{equation}

It is important to note that Hamiltonian vector fields need not be 
complete, that is, their flow need not exist for all flow parameters 
$t\in\reals$. For example, consider $P=\reals^{2}$ in canonical 
coordinates. The flow map $\reals\times P\rightarrow P$ is then given 
by $(t,(\qinit,\pinit))\mapsto (q(t;\qinit,\pinit),p(t;\qinit,\pinit))$, 
where the functions on the right hand side follow through integration of 
$X_f=\dot q(t)\partial_q+\dot p(t)\partial_p$, i.e.
\begin{equation}
\label{eq:symp-11}
\dot q(t)=(\partial_pf)(q(t),p(t))
\quad\mbox{and}\quad
\dot p(t)=-(\partial_qf)(q(t),p(t))\,,
\end{equation}
with initial conditions $q(0)=\qinit$, $p(0)=\pinit$.
As simple exercises one readily solves for the flows of $f(q,p)=h(q)$, 
$f(q,p)=h(p)$, where $h:P\rightarrow\reals$ is some $C^1$-function, 
or for the flow of $f(q,p)=qp$. All these are complete. But already 
for $f(q,p)=q^2p$ we obtain 
\begin{equation}
\label{eq:symp-12}
q(t;\qinit,\pinit)=\frac{\qinit}{1-\qinit t}
\quad\mbox{and}\quad
p(t;\qinit,\pinit)=\pinit\,(1-\qinit t)^2\,,
\end{equation}
which (starting from $t=0$) exists only for $t<1/\qinit$ when $\qinit>0$
and only for  $t>1/\qinit$ when $\qinit<0$.

\newpage
%%%%% APPENDIX 2

\section*{Appendix 2: The Lie algebra of $sl(2,\reals)$ and the 
absence of non-trivial, finite-dimensional representations by 
anti-unitary matrices} 
\label{sec:appendix-1}
Let us first recall the definition of a Lie algebra:
\begin{definition}
A \emph{Lie algebra over $\field$} (here standing for $\reals$ or
$\complex$) is a vector-space, $L$, over $\field$ together 
with a map $V\times V\rightarrow V$, called \emph{Lie bracket}
and denoted by $[\cdot,\cdot]$, such that the following conditions 
hold for all $X,Y,Z\in L$ and $a\in\field$:
\begin{alignat}{2}
\label{eq:app1:Lie1}
&[X,Y]=-[Y,X] && \mbox{antisymmetry}\,,\\
\label{eq:app1:Lie2}
&[X,Y+aZ]=[X,Y]+a[X,Z] &&\mbox{linearity}\,,\\
\label{eq:app1:Lie3}
&[X,[Y,Z]]+[Y,[Z,X]]+[Z,[X,Y]]=0\qquad &&\mbox{Jacobi identity}\,.
\end{alignat}
\end{definition}

Note that (\ref{eq:app1:Lie1}) and (\ref{eq:app1:Lie2}) together 
imply linearity also in the first entry. Any associative algebra
(with multiplication `$\cdot$') is automatically a Lie algebra 
by defining the Lie bracket to be the commutator 
$[X,Y]:=X\cdot Y-Y\cdot X$ (associativity then implies the Jacobi 
identity). Important examples are Lie algebras of square matrices,
whose associative product is just matrix multiplication.   
  
A sub vector-space $L'\subseteq L$ is a \emph{sub Lie-algebra}, 
iff $[X,Y]\in L'$ for all $X,Y\in L'$. A sub Lie-algebra is an 
\emph{ideal}, iff $[X,Y]\in L'$ for all $X\in L'$ and all 
$Y\in L$ (sic!). Two ideals always exist: $L$ itself and $\{0\}$; 
they are called the trivial ideals. A Lie algebra is called 
\emph{simple}, iff it contains only the trivial ideals. A map 
$\phi:L\rightarrow L'$ between Lie algebras is a \emph{Lie homomorphism}, 
iff it is linear and satisfies $\phi([X,Y])=[\phi(X),\phi(Y)]$ 
for all $X,Y\in L$. Note that we committed some abuse of notation 
by denoting the (different) Lie brackets in $L$ and $L'$ by the 
same symbol $[\cdot,\cdot]$. The kernel of a Lie homomorphism $\phi$
is defined by $\mbox{kernel}(\phi):=\{X\in L\mid \phi(X)=0\}$ and 
obviously an ideal in $L$.  

The Lie algebra denoted by $sl(2,\field)$ is defined by the vector 
space of traceless $2\times 2$ - matrices with entries in $\field$. 
A basis is given by 
\begin{equation}
\label{eq:app1:basis-sl2}
H=
\left(
\begin{array}{cr}
1&0\\
0&-1
\end{array}\right)\,,\quad
E_+=
\left(
\begin{array}{cc}
0&1\\
0&0
\end{array}\right)\,,\quad
E_-=
\left(
\begin{array}{cc}
0&0\\
1&0
\end{array}\right)\,.
\end{equation}
Its commutation relations are:
\begin{alignat}{2}
\label{eq:app1:sl2-commutators-1}
&[H,E_+]&&=\,2E_+\,,\\
\label{eq:app1:sl2-commutators-2}
&[H,E_-]&&=\,-2E_-\,,\\
\label{eq:app1:sl2-commutators-3}
&[E_+,E_-]&&\,=H\,.
\end{alignat}

The first thing we prove is that $sl(2,\field)$ is simple.
For this, suppose $X=aE_++bE_-+cH$ is a member of an ideal 
$I\subseteq sl(2,\field)$. From 
(\ref{eq:app1:sl2-commutators-1}-\ref{eq:app1:sl2-commutators-3})
we calculate
\begin{alignat}{1}
\label{eq:app1:sl2-simple1}
[E_+,[E_+,X]]&\,=\,-2b E_+\,,\\
\label{eq:app1:sl2-simple2}
[E_-,[E_-,X]]&\,=\,-2a E_-\,.
\end{alignat}  
Suppose first $b\not =0$, then (\ref{eq:app1:sl2-simple1}) shows that 
$E_+\in I$. Then (\ref{eq:app1:sl2-commutators-3}) implies $H\in I$,
which in turn implies through (\ref{eq:app1:sl2-commutators-2}) that 
$E_-\in I$ and hence that $I=L$. Similarly one concludes for $a\not =0$ 
that $I=L$. Finally assume $a=b=0$ and $c\not =0$ so that $H\in I$. Then   
(\ref{eq:app1:sl2-commutators-1}) and  (\ref{eq:app1:sl2-commutators-2})
show that $E_+$ and $E_-$ are in $I$, so again $I=L$.  
Hence we have shown that $I=L$ or $I=\{0\}$ are the only ideals.

Next consider the Lie algebra $u(n)$ of anti-Hermitean $n\times n$ 
matrices. It is the Lie algebra of the group $U(n)$ of unitary $n\times n$ 
matrices. If the group $SL(2,\field)$ had a finite-dimensional unitary 
representation, i.e. if a group homomorphism 
$D:SL(2,\field)\rightarrow U(n)$ existed for some $n$, then we would 
also have a Lie homomorphism $D_*:sl(2,\field)\rightarrow u(n)$ by 
simply taking the derivative of the map $D$ at $e$ (= identity of 
$SL(2,\reals)$). We will now show that, for any integer $n\geq 1$, 
any Lie homomorphism $\phi:sl(2,\field)\rightarrow u(n)$ is necessarily 
the constant map onto $0\in u(n)$. In other words, non-trivial Lie 
homomorphism from $sl(2,\field)$ to $u(n)$ do not exist. On the level 
of groups this implies that non-trivial (i.e. not mapping everything into 
the identity), finite dimensional, unitary representations of 
$SL(2,\field)$ do not exist. Note that for $\field=\reals$ and 
$\field=\complex$  these are (the double covers of) the proper 
orthochronous Lorentz groups in 2+1 and 3+1 dimensions respectively. 

To prove this result, assume $T:sl(2,\field)\rightarrow u(n)$ is a 
Lie homomorphism. To save notation we write $T(H)=:A$ and 
$T(E_{\pm})=:B_{\pm}$. Since $T$ is a Lie homomorphism we have 
$[A,B_+]=2B_+$, which implies 
\begin{equation}
\label{eq:app1:sl2-norep-1}
\mbox{trace}(B_+^2)\,=\, \tfrac{1}{2}\mbox{trace}
\left(B_+\,(AB_+ - B_+A)\right)=0\,,
\end{equation}
where in the last step we used the cyclic property of the trace. 
But $B_+$ is anti Hermitean, hence diagonalisable with 
purely imaginary eigenvalues $\{\i\lambda_1,\cdots,\i\lambda_n\}$
with $\lambda_i\in\reals$. The trace on the left side of 
(\ref{eq:app1:sl2-norep-1}) is then $-\sum_i\lambda_i^2$, which is 
zero iff $\lambda_i=0$ for all $i$, i.e. iff $B_+=0$. 
Hence $E_+\in \mbox{kernel}(T)$, which in turn implies 
$\mbox{kernel}(T)=sl(2,\field)$ since the 
kernel---being an ideal---is either $\{0\}$ or all of $sl(2,\field)$
by simplicity. This proves the claim, which is stated as 
Lemma\,\ref{lemma:sl2} of the main text


\begin{thebibliography}{10}

\bibitem{Abraham-Marsden:1978}
Ralph Abraham and Jerrold~E. Marsden.
\newblock {\em Foundations of Mechanics}.
\newblock The Benjamin/Cummings Publishing Company, Reading, Massachusetts,
  second edition, 1978.
%
\bibitem{Bordemann-etal:2000}
Martin Bordemann, Hans-Christian Herbig, and Stefan Waldmann.
\newblock {BRST} cohomology and phase space reduction in deformation
  quantisation.
\newblock {\em Communications of Mathematical Physics}, 210:107--144, 2000.
\newblock math.QA/9901015.
%
\bibitem{Dirac:1958}
Paul Dirac.
\newblock {\em The Principles of Quantum Mechanics}.
\newblock The International Series of Monographs in Physics 27. Oxford
  University Press, Oxford (UK), fourth edition, 1958.
\newblock 1982 reprint of fourth edition.

\bibitem{Dirac:1964}
Paul Dirac.
\newblock {\em Lectures on Quantum Mechanics}.
\newblock Belfer Graduate School of Science Monographs Series, Number Two.
  Yeshiva University, New York, 1964.

\bibitem{Gotay-1980}
Mark Gotay.
\newblock Functorial geometric quantization and {van Hove's} theorem.
\newblock {\em International Journal of Theoretical Physics}, 19(2):139--161,
  1980.

\bibitem{Gotay:1996-2torus}
Mark Gotay.
\newblock On a full quantization of the torus.
\newblock In J.-P. Antoine et~al., editors, {\em Quantization, Coherent States,
  and Complex Structure}, pages 55--62, New York, 1995. Bia{\l}owie\v{z}a,
  1994, Plenum Press.
\newblock math-ph/9507005.

\bibitem{Gotay:1996-2sphere}
Mark Gotay.
\newblock A {Groenewold - van Howe} theorem for {$S^2$}.
\newblock {\em Transactions of the American Mathematical Society},
  348:1579--1597, 1996.
\newblock math-ph/9502008.

\bibitem{Gotay:1999}
Mark Gotay.
\newblock On the {Groenewold - van Howe} problem for {$\mathbb{R}^{2n}$}.
\newblock {\em Journal of Mathematical Physics}, 40(4):2107--2116, 1999.
\newblock math-ph/9809015.

\bibitem{Groenewold:1946}
Hip Groenewold.
\newblock On the principles of elementary quantum mechanics.
\newblock {\em Physica}, 12:405--460, 1946.

\bibitem{Haag:1996}
Rudolf Haag.
\newblock {\em Local Quantum Physics: Fields, Particles, Algebras}.
\newblock Texts and Monographs in Physics. Springer Verlag, Berlin, second
  edition, 1996.

\bibitem{Henneaux-Teitelboim:1992}
Marc Henneaux and Claudio Teitelboim.
\newblock {\em Quantization of Gauge Systems}.
\newblock Princeton University Press, Princeton, USA, first edition, 1992.

\bibitem{Isham-LesHouches:1984}
Chris Isham.
\newblock Topological and global aspects of quantum theory.
\newblock In B.S. DeWitt and R.~Stora, editors, {\em Relativity, Groups and
  Topology II}, pages 1059--1290, Amsterdam, 1984. Les Houches 1983, Session
  XL, North-Holland Physics Publishing.

\bibitem{Joos-etal:2003}
Erich Joos, {H.-Dieter} Zeh, Claus Kiefer, Domenico Giulini, Joachim Kupsch,
  and Ion-Olimpiu Stamatescu.
\newblock {\em Decoherence and the Appearence of a Classical World in Quantum
  Theory}.
\newblock Springer Verlag, Berlin, second edition, 2003.

\bibitem{Loll:2003}
Renate Loll.
\newblock A discrete history of the {Lorentzian} path integral.
\newblock In D.~Giulini, C.~Kiefer, and C.~L\"ammerzahl, editors, {\em Aspects
  of Quantum Gravity - From Theory to Experimental Search}, Berlin, 2003.
  Springer Verlag.
\newblock E-Archive: hep-th/0212340.

\bibitem{Mohaupt:2003}
Thomas Mohaupt.
\newblock Introduction to string theory.
\newblock In D.~Giulini, C.~Kiefer, and C.~L\"ammerzahl, editors, {\em Aspects
  of Quantum Gravity - From Theory to Experimental Search}, Berlin, 2003.
  Springer Verlag.
\newblock E-Archive: hep-th/0207249.

\bibitem{Reed-Simon-1:1972}
Mike Reed and Barry Simon.
\newblock {\em Functional Analysis}.
\newblock Methods of Modern Mathematical Physics. Academic Press, New York,
  first edition, 1972.

\bibitem{Thiemann:2003}
Thomas Thiemann.
\newblock Lectures on loop quantum gravity.
\newblock In D.~Giulini, C.~Kiefer, and C.~L\"ammerzahl, editors, {\em Aspects
  of Quantum Gravity - From Theory to Experimental Search}, Berlin, 2003.
  Springer Verlag.
\newblock E-Archive: gr-qc/0210094.

\bibitem{VanHowe:1951-b}
L\'eon van Howe.
\newblock Sur certaines repr\'esentations unitaires d'un groupe infini de
  transformations.
\newblock {\em Memoirs de l'Academie Royale de Belgique (Classe des Sciences)},
  26(6):61--102, 1951.

\bibitem{VanHowe:1951-a}
L\'eon van Howe.
\newblock Sur le probl\`eme des relations entre les transformations unitaires
  de la m\'ecanique quantique et les transformations canoniques de la
  m\'ecanique classique.
\newblock {\em Academie Royale de Belgique Bulletin de la Classe des Sciences},
  37(5):610--620, 1951.
%
\end{thebibliography}
\end{document}